\documentstyle[prl,aps,preprint,eqsecnum,tighten,epsf]{revtex}
\begin{document}
\title{On the Finite Size Scaling in Disordered Systems}
\author{H. Chamati$^1$, E. Korutcheva$^2$\thanks{Permanent address:
G. Nadjakov Inst. Solid State Physics, Bulgarian Academy of
Sciences, 1784 Sofia, Bulgaria and Regular Associate Member of ICTP,
Trieste, Italy}, N.S. Tonchev$^1$\thanks{Senior
Associate Member of ICTP, Trieste, Italy. e-mail:
tonchev@issp.bas.bg}}
\address{
$^1$G.Nadjakov Institute of Solid State Physics, Bulgarian
Academy of Sciences,\\
1784 Sofia, Bulgaria\\
$^2$Dep. F\'{\i}sica Fundamental, Universidad Nacional de
Educaci\'on a Distancia,\\
c/ Senda del Rey No 9 ,  28080 Madrid, Spain}
\maketitle
\draft
\begin{abstract}
The critical behavior of a  quenched random hypercubic sample of
linear size $L$ is considered, within the
``random-$T_{c}$'' field-theoretical
model, by using the renormalization group method. A
finite-size scaling behavior is established and analyzed near the
upper critical dimension $d=4-\epsilon$ and some universal
results are obtained. The problem of self-averaging is clarified
for different critical regimes.
\end{abstract}

\pacs{PACS Numbers: 75.10.Nr, 05.70Fh, 64.60.Ak, 75.40.Mg}

\section{Introduction}
The description of effects of disorder on the critical behavior of
finite-size systems have attracted a lot of interest
\cite{WD1,ABH,PSZ,WeD,ABW,BFMMPR,MG,BPB}. Up to now a
discussion takes place of whether the introduced disorder
influences the finite-size scaling (FSS) results\cite{PSZ,BPB},
compared to the standard FSS results, known for pure systems
\cite{Barber,Privman,BDT}. A formulation of general
FSS concepts for the case of disorder is strongly complicated due
to the additional averaging over the different random samples. For
a random sample with volume $L^{d}$, where $L$ is a linear
dimension, any observable property $X$,
singular in the thermodynamic limit, has different
values for different realizations of the randomness and can be
considered as stochastic variable with mean $\overline{X}$ and
variance $(\Delta X)^{2}:=\overline{X^{2}} -\overline{X}^{2}$,
where the over line indicates an average over all realizations of
the randomness. Here, an important theoretical problem of interest
is related with the property of self-averaging (SA) \cite{BY}. If
the system does not exhibit SA a measurement performed on a single
sample does not give a meaningful result and must be repeated on
many samples. A numerical study of such a system also will be
quite difficult. This point has been studied recently by means of
FSS arguments \cite{WD1,WeD}, renormalization group (RG) analysis
\cite{ABH,ABW} and Monte Carlo simulations \cite{WeD,BFMMPR}. The
quantity under inspection is the relative variance
$R_{X}(L):=(\Delta X)^{2}/\overline{X}^{2}$. A system is said to
exhibit ``strong SA'' if $R_{X}(L)\sim L^{-d}$ as $L\to\infty$.
This is the case if the system is away from criticality, i.e. if
$L\gg\xi$. At the criticality, i.e. when $L\ll\xi$, the situation
depends whether the randomness is irrelevant
($\nu_{pure}>2/d$, e.g. ``pure'', P-case) or relevant ($\nu_{random}>2/d$,
e.g. ``random'' R-case) \cite{ABH}. One calls the former case
``weak SA'', since $R_{X}(L)\sim L^{(\alpha/\nu)_{pure}}$, and the
latter case ``no SA'', since $R_{X}(L)$ is fixed nonzero universal
quantity even in the thermodynamic limit \cite{ABH}.

The lack of SA in disordered systems implies that the standard FSS
breaks down. Thus, one needs to formulate a FSS theory, suitable for
the case under consideration. There is an ongoing activity in this
field but an understanding of the problems at a deeper level is
still desirable \cite{WD1,PSZ,BPB}.

The successful application of the field theoretical methods
\cite{BZJ,RGJ} for the analytic calculations of the size-dependent
universal scaling functions for pure systems makes the possible
extension of these methods very appealing also in the case of
disordered systems. The appropriate field theory is the
N-component $\psi^{4}$ theory with a "random- $T_{c}$" term
\cite{Kh,L,GL,JK,Sh,Jug,KU}. The analytic difficulties raised by the
disorder usually are avoided with the replica trick
by solving an effective pure problem \cite{GL}. However, the perturbative
structure of the theory is still much more complicated than for
the corresponding pure system, arising additional difficulties in
the applicability of the ideas proposed in \cite{BZJ,RGJ}.
Some of them are of rather hard computational nature.

Between the
generic models for magnetism, the most popular and relatively well
studied by means of Monte Carlo simulations is the Ising model (i.e.
$N=1$) due to the fact that in three dimensions it is the model for
which in accordance with the Harris criterion, randomness is relevant.
But its RG analysis is complicated as a result of the well known
accidental degeneracy in the recursion relations that needs higher
order in $\epsilon$ and so the use of the loop expansion to second
order (see e.g.\cite{Kh,JK,Sh,Jug}). This leads to the apparent
computational difficulties in the finite-size treatment of the system.
The $\epsilon$- calculations based on the loop expansion to second
order are not done even for the pure finite-size systems. In this
situation, more attractive for the finite-size RG study are the
disordered XY ($N=2$) and Heisenberg ($N=3$) cases as simpler and
having the same qualitative features governed in $d=4-\epsilon$
dimensions by a random fixed point.

Other problems have more basic nature and are related with the
breaking of the replica symmetry (see \cite{MGT}). Since its
deeper understanding is still lacking, they are beyond of our
interest in the present study.

In this paper we analyze  the finite-size properties of an
N-component ($N>1$) model of randomly diluted magnet
with hypercubic geometry of linear size $L$. Exact
calculations are performed in the mean-field regime $d>4$, and up
to the first order in $\epsilon$ near the upper critical dimension
$d_{c}=4$. Although in this case, the problem of usefulness of the
corresponding series expansions away from the dimension
$d=4-\epsilon$ arises and is questionable (see \cite{PV} and refs.
therein), we shall show
that many generic FSS properties of the model can be established
and we hope this would have implications for more realistic cases.

The paper is organized as follow. In Sections \ref{secmodel} and
\ref{seceff} we define the model and the effective Hamiltonian. In
Section \ref{seczeromode} we perform the  analysis in the zero-mode
approximation. In Sections \ref{secshift} and \ref{seccoupling} we give
the expressions for the shift of the critical temperature and the
renormalized coupling constants in first order in $\epsilon$. Section
\ref{secfss} deals with the verification of the FSS and the analysis of
the problem of Self-Averaging is given in Section \ref{secsa}. Finally in
Section \ref{discussion} we present our main conclusions.

\section{Model}\label{secmodel}

We consider the "random - $T_{c}$" Ginzburg-Landau-Wilson model of
disordered ferromagnets (see, e.g. \cite{Kh,L,GL,JK,Sh,Jug})
\begin{equation}
\label{eqGLW} {\cal H}_{r}=-\frac{1}{2} \int_{L^{d}} d^{d}x [ t |
\psi({\bbox x})|^{2} + \varphi({\bbox x}) |{\bbox \psi}({\bbox
x})|^{2} +c|\nabla \psi ({\bbox x})|^{2} + \frac{u}{12} |
\psi({\bbox x})|^{4}],
\end{equation}
where $ \psi({\bbox x})$ is a N-component field with $\psi^2({\bbox
x})=\sum_{i=1}^N\psi_i^2({\bbox x})$ and the random variable
$\varphi({\bbox x})$ has a Gaussian distribution
\begin{equation}
\label{eqGaus}
 P(\varphi({\bbox x}))=\frac{\exp[-\frac{\varphi({\bbox x})^{2}}
{2\Delta}]}{\sqrt{2\pi\Delta }}
\end{equation}
with mean
\begin{equation}
\overline{\varphi({\bbox x})} = 0
\end{equation}
and variance
\begin{equation}
\label{eqGU}
\overline{\varphi({\bbox x})\varphi({\bbox x} \prime)}=
\Delta \delta^{d}({\bbox x} - {\bbox x}').
\end{equation}
The over line in (\ref{eqGU}) indicates a random average performed
with the distribution $P(\varphi({\bbox x}))$.
Here we will consider a system in a
finite cube of volume $L^{d}$ with periodic boundary conditions.
This means that the following expansion takes place
\begin{equation}\label{fpbc}
\psi({\bbox x})=\frac{1}{L^d}\sum_{{\bbox k}} \tilde{\psi} ({\bbox
k})\exp(i{\bbox k}.{\bbox x})
\end{equation}
and
\begin{equation}
\label{fp}
 \varphi({\bbox x})=\frac{1}{L^d}\sum_{{\bbox k}}
 \tilde{\varphi}({\bbox k})\exp(i{\bbox k}.{\bbox x}),
\end{equation}
where ${\bbox k}$ is a discrete vector with components $k_{i}=2\pi
n_{i}/L$ , $n_{i}=0,\pm 1,\pm 2,...$, $i=1,...,d$ and a cutoff
$\Lambda \sim a^{-1}$ ($a$ is the lattice spacing). In this paper, we
are interested in the continuum limit, i.e. $a \to 0$.

In our case of quenched randomness one must average the logarithm of
the partition function over the Gaussian distribution (\ref{eqGaus}) to
produce the free energy
\begin{equation}
\label{frGLW}
{\cal F}\left[{\cal H}_{r}\right]=
-\int_{-\infty}^{\infty}D\varphi({\bbox x})P(\varphi({\bbox x}))
\ln {\cal Z}_{r},
\end{equation}
where
\begin{equation}
\label{zNGLW}
{\cal Z}_{r}=Tr_{\psi}\exp[{\cal H}_{r}].
\end{equation}

It is well known that the direct average of ${\cal H}_{r}$
over the Gaussian
leads to equivalent results \cite{eqv} for the critical behavior as the
$n=0$ limit of the following "pure" translationally invariant model
\cite{footnote}
\begin{eqnarray}
\label{GLWcub}
{\cal H}_{p}(n)&=& - \frac{1}{2} \sum_{\alpha=1}^{n} \int_{L^{d}}
d^{d}x [ t |
\psi_{\alpha}({\bbox x})|^{2} +c|\nabla{\psi}_{\alpha}({\bbox x})|^{2}
+ \frac{u}{12} | \psi_ {\alpha}({\bbox x})|^{4}]
\nonumber \\
&& +\frac{\Delta}{8} \sum_{\alpha, \beta=1}^{n} \int_{L^{d}}
d^{d}x | \psi_{\alpha}({\bbox x})|^{2} | \psi_{\beta}({\bbox
x})|^{2} ] .
\end{eqnarray}
Here $\psi_{\alpha}({\bbox x})$, $\alpha=1,\ldots,n$ ($n$ beings the
number of replicas) are components of an $(n\times N)$-components
field $\vec\psi({\bbox x})$. Because of this equivalence, the
model ${\cal H}_{p}$ has been the object of intensive
field-theoretical studies (see \cite{PV} and refs. therein) in the
bulk case. Much less is known for the equivalence of ${\cal
H}_{r}$ and the $n=0$-limit of ${\cal H}_{p}$ in the finite - size
case. Problems may arise when finite-size techniques are used,
since both procedures $L\to\infty$ and removing of disorder by the
``trick'' $n \to 0$ may not commute.

\section{The effective Hamiltonian}\label{seceff}
In this work we will use the RG technique introduced in \cite{BZJ}
and \cite{RGJ} for studying pure systems with finite geometry.
This technique permits explicit analytical calculations above and
in the neighborhood of the upper critical dimension. The main idea
is to expand the field in (\ref{eqGLW}) in Fourier modes and then
to treat the zero mode separately from the nonzero modes. The
nonzero modes can be treated by the methods developed for the bulk
systems (e.g. loop expansion), while the zero mode, whose
fluctuations are damped at the critical temperature,
has to be treated exactly. This overcomes the problems due to
the infra red (IR) divergences that take place in finite size
systems (see reference \cite{brezin82}).

In our more complicated case we have two possibilities: to
consider the random model Eq.~(\ref{eqGLW}) or to consider the
replicated pure model Eq. (\ref{GLWcub}). The last one is closer
to the case treated in \cite{BZJ} and \cite{RGJ} by getting around
the difficulties due to random average performed with
$P(\varphi({\bbox x}))$ and is used in the present study.
For this case, the replicated partition function is
given by
\begin{equation}
{\cal Z}_{p}(n)=\int{\cal D}\psi\exp\left[{\cal H}_p(n)\right].
\end{equation}
We decompose the field $\psi({\bbox x})$ into a zero
momentum component $\phi=L^{-d}\int d^{d}x\psi({\bbox x})$, which
plays the role of the uniform magnetization and a second part depending
upon the non-zero modes $\sigma=L^{-d}\sum_{\bbox k \neq
0}\tilde{\psi}(\bbox k)\exp(-i{\bbox k}.{\bbox x})$. After some algebra,
the partition function can be expressed as
\begin{eqnarray}\label{effpart}
{\cal Z}_{p}(n)&=&\int{\cal D}\phi{\cal D}\sigma\exp\left\{
-\frac{L^{d}}2\sum_{\alpha=1}^n\left(r_0\phi^2_{\alpha}+
\frac{u_0}{12}\phi^4_{\alpha}\right)+\frac{L^d\Delta_0}8
\left(\sum_{\alpha=1}^n\phi^2_{\alpha}\right)^2\right.\nonumber\\
&&\left.-\frac12\sum_{\alpha=1}^n{\sum_{\bbox k}} \left[r_0+{\bbox
k}^2+\frac{u_0}2\phi_\alpha^2-\frac{\Delta_0}2
\sum_{\beta=1}^n\phi^2_{\beta}\right]
\sigma_{\alpha}^2+{\rm higher\ powers\ of\ \sigma}\right\}.
\end{eqnarray}
Here the terms involving $\int d^dx\sigma$ vanishes since $\sigma$
depends only on non zero modes.
The terms containing $\sigma$ are treated using diagram expansion,
leading to the effective Hamiltonian in the one-loop approximation
\begin{equation}\label{effectiveHam}
{\cal H}^{\rm eff}_{p}(n)=-\frac{L^{d}}{2}\sum_{\alpha=1}^n\left(\tilde
t(n)\phi^2_{\alpha}+\frac{\tilde u (n)}{12}\phi^4_{\alpha}\right)
+\frac{L^d\tilde\Delta
(n)}8\left(\sum_{\alpha=1}^n\phi^2_{\alpha}\right)^2,
\end{equation}
where ${\tilde t}(n), {\tilde u}(n) $ and ${\tilde \Delta}(n)$ will be
presented bellow.
With the help of the identity
\begin{equation}
\exp\left(\frac{aA^{2}}{2}\right)=\frac{1}{(2\pi
a)^{1/2}}\int_{-\infty}^{\infty}dy\exp[-(1/2a)y^{2}+yA] ,
\end{equation}
we get
\begin{equation} \label{ihs}
{\cal Z}_{p}^{\rm eff}(n)=Tr_{\phi} \exp[{\cal H}_{p}^{\rm eff}(n)]=
\int_{-\infty}^{\infty}dy P_{n}\left(y\right)
\left[S_{N}\int_{0}^{\infty}
d|\phi||\phi|^{N-1}\exp({\cal H}_{r}^{\rm eff}(n))\right]^{n},
\end{equation}
where
\begin{equation}\label{reff}
{\cal H}_{r}^{\rm
eff}(n)=-\frac{1}{2}L^{d}\left [\left({\tilde
t(n)}+ \frac{y}{L^{d/2}}\right) |\phi|^{2}+\frac{1}{12}{\tilde
u(n)} |\phi|^{4} \right]
\end{equation}
is an effective Hamiltonian with a random variable $y$ with
Gaussian distribution (depending on ${\tilde \Delta}(n)$)
\begin{equation}\label{eqGn}
P_{n}(y)=\frac{\exp\left(-\frac{y^{2}} {2{\tilde
\Delta}(n)}\right)}{\sqrt{2\pi{\tilde \Delta}(n) }}
\end{equation}
and $S_{N}=\frac{2\pi^{N/2}}{\Gamma(N/2)}$ is the surface of a
$N$-dimensional unit sphere.

Let us note that the above mentioned equivalence between the models
(\ref{eqGLW}) and (\ref{GLWcub}) in the used approximation
may be mathematically expressed, within the used approximation
by the following relation:
\begin{equation}\label{meeq}
{\cal F}\left[{\cal H}_{r}\right]=-\frac{\partial}{\partial
n} {\cal Z}_{p}(n)\mid_{n=0}.
\end{equation}
From Eqs. (\ref{ihs}) and (\ref{meeq}), and by using the identity
\begin{equation}
\frac{\partial}{\partial n}A^{n}(n)\mid_{n=0}
=\ln A(0)
\end{equation}
for the free energy, we get
\begin{equation}\label{A1B2}
{\cal F}\left[{\cal H}_{r}\right]=-\int_{-\infty}^{\infty}dy
P_{0}(y)\ln {\cal Z}_{r}(0),
\end{equation}
where
\begin{equation}
\label{zet0}
{\cal Z}_{r}(0)=S_{N}\int_{0}^{\infty}
d|\phi||\phi|^{N-1}\exp[{\cal H}_{r}^{\rm eff}(0)]
\end{equation}
is the partition function for the random system (\ref{reff})
after taking the limit $n\to0$. The
obtained effective ``random-$T_{c}$ model'' (\ref{reff}), 
distributed with Gaussian weight (\ref{eqGn}), is the analytic
basis of this paper. The effective constants ${\tilde t}(n)$,
${\tilde u}(n)$ and ${\tilde \Delta}(n)$ involve $n$ and
finite-size $L$ as parameters. For describing the finite-size
properties of the initial model (\ref{eqGLW}), as follows from Eqs.
(\ref{A1B2}) and (\ref{zet0}), it is necessary to set $n$ to zero.
In the next sections we shall consider the results of this
procedure.

\section{The FSS expression for the free energy and cumulants in the
zero mode approximation}\label{seczeromode}

If we neglect the loop corrections this corresponds to the
mean-field approximation. Then the zero mode playing the role of
the uniform magnetization may be treated exactly. In this case the
effective Hamiltonian (\ref{effectiveHam}) of the model reduces to
\begin{equation}\label{GLWcubL}
{\cal H}^{MF}_{p}=-\frac{1}{2}L^{d}
\left[t\sum_{\alpha=1}^n\phi_\alpha^{2} +
\frac{1}{12}u \sum_{\alpha=1}^{n} ( \phi_{\alpha}^{2})^{2}-
\frac{\Delta}{4}\left(\sum_{\alpha=1}^n\phi_\alpha^{2}\right)^{2}\right].
\end{equation}

Now using an appropriate rescaling of the field $ |\phi|=
(uL^{d})^{-1/4}\Phi$ and introducing the scaling variable
\begin{equation}\label{ispr}
\mu=tL^{d/2}u^{-1/2},
\end{equation}
for the partition function Eq. (\ref{zet0}) in the mean-field
approximation we obtain
\begin{equation}
\label{alabala}
{\cal Z}_{r}^{MF}(0)=(uL^{d})^{-N/4}
{\cal I}_{N}(\mu+y/u^{1/2}),
\end{equation}
where we have introduced the following auxiliary function
\begin{equation}
{\cal I}_{N}(z)=S_N
\int_{0}^{\infty}d\Phi \Phi^{(N-1)}\exp\left\{-\frac12\left[z
\Phi^{2}+\frac1{12}\Phi^{4}\right]\right\}.
\end{equation}

From Eqs. (\ref{A1B2}) and (\ref{alabala}) we get for the free energy
\begin{equation}\label{zbala}
{\cal F}\left[{\cal H}_{r}^{MF}\right]=-\frac{1}
{\sqrt{2\pi\Delta}}\int_{-\infty}^{\infty}dy
\exp\left(-\frac{1}{2}\frac{y^{2}}{\Delta}\right)
\ln\left[(uL^d)^{-N/4}{\cal I}_{N}\left(\mu+y/u^{1/2}\right)\right].
\end{equation}
If we introduce a second scaling variable
\begin{equation}\label{svfe}
\lambda=\frac\Delta u,
\end{equation}
and using Eqs. (\ref{NN}) (see appendix), Eq. (\ref{zbala})
takes its final form
\begin{equation}\label{scba}
{\cal F}\left[{\cal H}_{r}^{MF}\right]=
-\frac{1}{\sqrt{2\pi\lambda}}\int_{-\infty}^{\infty}dx
e^{-(x-\mu)^{2}/(2\lambda)}
\ln\left[D_{-N/2}(\sqrt{3}x)\right]-\frac34(\lambda+\mu^2)
+\frac N4\ln\left(\frac{uL^{d}}{12\pi^2}\right).
\end{equation}
For the Ising case $N=1$ a similar expression for the
quenched free energy
in a slightly different context is obtained and its analytic structure
is studied in references \cite{BMM,McK,AML}. Obviously
(\ref{scba}) is well defined for any positive $\lambda$ and in the
limit $\lambda \to 0$ we recover the well known result for the
free energy of the pure model.

In addition to the free energy, one also needs to know the correlation
functions. Within the replica method the averages of the fields
$\{\phi_{\beta}\}$ are defined by (see e.g. \cite{BY})
\begin{equation}\label{c2p}
\overline{\langle|\phi_{\beta}|^{2m}\rangle}_{{\cal H}_{r}^{MF}}=
\lim_{n \to 0}\left[{\cal Z}^{MF}_{p}(n)^{-1}S_{N}^n 
\int\left(\prod_{\alpha=1}^
{n}d |\phi_{\alpha}|\right)(|\phi_{\alpha}|)^{N-1}
(|\phi_{\beta}|)^{2m} \exp({\cal H}_{p}^{MF})\right],
\end{equation}
where
\begin{equation}\label{laba}
{\cal Z}_{p}^{MF}(n)=S_{N}^n 
\int\left(\prod_{\alpha=1}^
{n}d |\phi_{\alpha}|\right)(|\phi_{\alpha}|)^{N-1}
\exp({\cal H}_{p}^{MF}).
\end{equation}
Note that the final result must be independent of replica index
$\beta$, because ${\cal H}_{p}^{MF}$ is invariant under permutation of
the replicas. After taking the limit $n \rightarrow 0$, we end with
the following expression:
\begin{equation}\label{cum1}
\overline{{\cal M}_{2m}}:=
\overline{ \langle |\phi_{\beta}|^{2m}\rangle}_{{\cal H}_{r}^{MF}} =
\frac{(uL^d)^{-m/2}}{\sqrt{2 \pi \lambda}} \int_{-\infty}^{\infty} dx
\frac{ {\cal I}_{N+2m}(x)}{{\cal I}_{N}(x)}
e^{-(x-\mu)^{2}/2\lambda}.
\end{equation}
In a similar way
\begin{equation}
\label{cum2}
\overline{({\cal M}_{2})^{2}}:=
\overline{ \langle
|\phi_{\alpha}|^{2} |\phi_{\beta}|^{2} \rangle}_{{\cal H}_{r}^{MF}} =
\frac{(uL^d)^{-1}}{\sqrt{2 \pi \lambda}} \int_{-\infty}^{\infty} dx
\left[\frac{ {\cal I}_{N+2}(x)}{{\cal I}_{N}(x)}\right]^{2}
e^{-(x-\mu)^{2}/2\lambda} .
\end{equation}
From Eqs. (\ref{cum1}) and (\ref{cum2}), when $\mu=0$ and $N=1$ we
obtain the results of Ref. \cite{BFMMPR} .

In terms of the normalized magnetization ${\cal M}$ the
susceptibility is given as
\begin{equation}\label{M2}
\chi=L^{d}\overline{{\cal M}_{2}}.
\end{equation}
Another quantities of importance for numerical analysis is the Binder
cumulant defined by
\begin{equation}\label{Bc}
B=1-\frac{1}{3}\frac{\overline{ {\cal M}_{4}}}
{\overline{{\cal M}_{2}}^{2}}
\end{equation}
and the cumulant, specific for the random system defined as
\begin{equation}\label{rnc}
R= \frac{\overline{({\cal M}_{2})^{2}}-
\overline{{\cal M}_{2}}^{2}}
{\overline{{\cal M}_{2}}^{2}}.
\end{equation}

Since the parameter $R$ is the relative variance of the
observable (the susceptibility), as we said in the Introduction, it is a
measure of the self-averaging in the random system.
If self-averaging takes place this quantity should be zero in the
thermodynamic limit.


\section{Finite-size shift of $T_{\lowercase{c}}$ : lowest order in $\epsilon$}
\label{secshift}

The perturbatively calculated parts of the free energy and cumulants,
which contain contributions of all nonzero modes, depend, to one-loop
order, on the shift of the critical temperature and on the
renormalized coupling constants $u$ and $\Delta$. The application of
the finite-size $\epsilon$-expansion to the model system
(\ref{GLWcub}) requires the corresponding renormalization constants.

To one loop order, using the minimal subtraction scheme,
before taking the $n \rightarrow 0$ limit, we obtain:
\begin{mathletters}\label{eqZZZg}
\begin{eqnarray}
Z_{t}&=&
1+\frac{N+2}{6\epsilon}\hat{u}-\frac{2+nN}{2\epsilon}\hat{\Delta},\\
Z_{u}&=&
1+\frac{N+8}{6\epsilon}\hat{u}-\frac{6\hat{\Delta}}{\epsilon},\\
Z_{\Delta}&=&
1+\frac{N+2}{3\epsilon}\hat{u}-\frac{8+nN}{2\epsilon}\hat{\Delta}.
\end{eqnarray}
\end{mathletters}
In Eqs. (\ref{eqZZZg}),
\begin{mathletters}\label{eqZu}
\begin{eqnarray}
\hat{u}&=& u\frac{2}{(4\pi)^{d/2}\Gamma(d/2)},\\
\hat{\Delta}&=& \Delta\frac{2}{(4\pi)^{d/2}\Gamma(d/2)}.
\end{eqnarray}
\end{mathletters}
The $\beta$ functions associated to $\hat{u}$ and $\hat{\Delta}$
have the form
\begin{mathletters}\label{bfc}
\begin{eqnarray}
\beta_{u}&=&-\hat{u}\epsilon + \frac{N+8}{6}\hat{u}^{2} -
6\hat{u}\hat{\Delta},\\
\beta_{\Delta}&=&-\hat{\Delta}\epsilon - \frac{8 +
nN}{2}\hat{\Delta}^{2} + \frac{N+2}{3}\hat{u}\Delta.
\end{eqnarray}
\end{mathletters}
The fixed points of this system first have been studied in
\cite{BGL} and for the purposes of the impurity problem in
\cite{GL}. The values of $\hat{u}$ and $\hat{\Delta}$ in the fixed
point, interesting in the random case, are
\begin{mathletters}\label{zhfp}
\begin{eqnarray}
\hat{u}^{\star}(n)&=&\frac{6(4-nN)}{16(N-1)-nN(N+8)}\epsilon,\\
\hat{\Delta}^{\star}(n)&=&\frac{2(4-N)}{16(N-1)-nN(N+8)}\epsilon.
\end{eqnarray}
\end{mathletters}
The corresponding expression for the exponent $\nu$ up to the first
order of $\epsilon$, is:
\begin{equation}
\label{nurn}
\frac{1}{\nu(n)}=2-\frac{6N(1-n)}{16(N-1)-nN(N+8)}\epsilon.
\end{equation}

It should be noted here that we shall consider $N$-component fields
with $1<N<N_{c}(d)$,
where $N_{c}(d)=4-4\epsilon +{\cal O}(\epsilon^{2})$  is the
critical number of spin components that defines the stability of
the random fixed point in the $n=0$ limit. The stability of the
different fixed points of the model has been also considered in
\cite{DG}. The analysis of the Ising case $(N=1)$ needs to perform
a loop expansion to second order (see Introduction) and is beyond
the scope of the present study.

\begin{figure}[tbp]
\vspace{1cm}
\epsfxsize=3.2in
\centerline{\epsffile{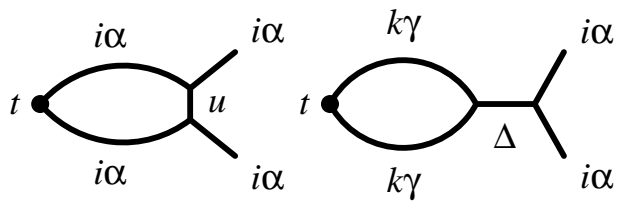}}
\vspace{0.2in}
\caption{One loop contributions to the reduced temperature $t$}
\label{fig1}
\end{figure}

As it was explained above, the loop corrections will be treated
perturbatively on the nonzero ${\bbox k}$ modes. In the lowest
order in $\epsilon$, this procedure generates a shift of the
critical temperature $t \to \tilde{t}(n)$;
\begin{equation}
\label{ttil} \tilde{t}(n)=tZ_{t}+t_{L} ,
\end{equation}
where the term $tZ_t$ is coming from the one-loop counterterm (see
(\ref{eqZZZg})), and
\begin{equation}
\label{eqnp} t_{L}=\left[\frac{N+2}{6}\hat{u}-
\frac{2+Nn}{2}\hat{\Delta}\right]\frac{1}{L^{d}}{\sum_k}'\frac{1}{\bbox
k^{2}+t}
\end{equation}
is the finite-size correction. The two diagrammatic contributions
for $t_{L}$ are shown in FIG. \ref{fig1}. Both diagrams from the
$u$ and the $\Delta$ contributions differ only by their numerical
factors. The prime in the $d$-fold sum in the above equations
denotes that the term with a zero summation index has been
omitted.

After some algebra (details  for the pure case $(\Delta=0)$ see in
\cite{BZJ}), near the upper critical dimension $d=4-\epsilon$, we
obtain
\begin{equation}\label{urtiq}
\tilde{t}(n)=t + \left[\frac{N+2}{12}\hat{u}-
\frac{2+Nn}{4}\hat{\Delta}\right]\left(
t\ln t +4L^{-2}F_{4,2}(tL^{2})\right),
\end{equation}
where
\begin{equation}\label{eqF}
F_{d,2}(x)=\int^{\infty}_{0}dz\exp\left(-\frac{xz}{(2\pi)^{2}}\right)
\left[\left(\sum_{\ell=-\infty}^{\infty}e^{-z\ell^{2}}\right)^{d}
-1-\left(\frac{\pi}{z}\right)^{d/2}\right],
\end{equation}
Some particular values of the constant $F_{d,2}(0)$ and a method
of calculation are given in \cite{CT}.

At the fixed point $\hat{u}=\hat{u}^{\star}(n)$, $\hat{\Delta}
=\hat{\Delta}^{\star}(n)$, up to the first order in $\epsilon$,
the terms proportional to $\ln L$ cancel and Eq. (\ref{urtiq}) can
be written in the following scaling form
\begin{equation}\label{svqst}
{\tilde t}(n)L^2=y-\frac{3(n-1)N}{16(N-1)-nN(N+8)}\left[y\ln y
+4F_{4,2}(y)\right]\epsilon,
\end{equation}
where the scaling variable $y=tL^{1/\nu(n)}$ has been introduced.

In the $n=0$ limit, the expression
for the exponent measuring the divergence of the correlation
length is \cite{L,GL}:
\begin{equation}
\label{nur}
\frac{1}{\nu_{R}}\equiv\frac{1}{\nu(0)}=2-\frac{3N}{8(N-1)}\epsilon,
\end{equation}
instead of the critical exponent for the pure case
\begin{equation}
\label{nup}
\frac{1}{\nu_{P}}=2-\frac{N+2}{N+8}\epsilon.
\end{equation}
The scaling form (\ref{svqst}) can be written for this case as
\begin{equation}
\label{epstf}
\tilde{t}(0)L^{2}=y + \frac{3N }{16(N-1)}\left[y\ln y +
4F_{4,2}(y)\right]\epsilon,
\end{equation}
where $y=tL^{1/\nu_{R}}$.

From the above expression one can obtain the large-L asymptotic
form of the $T_{c}(\infty)$ shift, i.e.
\begin{equation}\label{eqsh}
T_{c}(L)-T_{c}(\infty) \sim L^{-1/\nu_{R}}.
\end{equation}
In Eq. (\ref{eqsh}), $T_{c}(L):=\overline{T_{c}(\varphi,L)}$
denotes the average pseudo critical temperature
($T_{c}(\varphi,L)$ is pseudo critical temperature for a specific
random realization $\varphi(x)$) and $T_{c}(\infty)=\lim_{L \to
\infty}T_{c}(L)$. Eq. (\ref{eqsh}) was suggested in \cite{ABH}.
Combined with the fenomenological FSS theory it gave rise to the
lack of SA, and is confirmed by numerical studies (see
\cite{WeD}). Here it is verified independently and directly.


\section{Renormalization of the coupling constants : lowest order
in $\epsilon$}\label{seccoupling}

\begin{figure}[tbp]
\vspace{1cm}
\epsfxsize=4in
\centerline{\epsffile{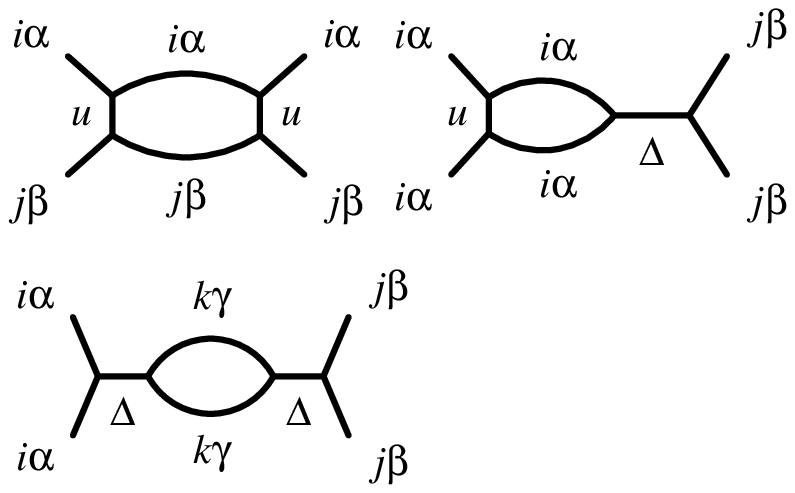}}
\vspace{0.2in}
\caption{One loop contributions to the couplings $u$ and $\Delta$}
\label{fig2}
\end{figure}

We perform the renormalization in a similar way also for
$\tilde{u}(n)$ and $\tilde{\Delta}(n)$ by taking into account the
diagrammatic contributions, from $u$ and $\Delta$ shown in FIG.
\ref{fig2}. The result is
\begin{mathletters}
\begin{eqnarray}
\tilde{u}(n)&=& uZ_u+u_{L}\\
\tilde{\Delta}(n)&=& \Delta Z_\Delta+\Delta_{L},
\end{eqnarray}
\end{mathletters}
where $uZ_{u}$ and $\Delta Z_{\Delta}$ are the one-loop
counterterms for the coupling constants, and
\begin{mathletters}\label{udre}
\begin{eqnarray}
u_{L}&=& -\left[u^{2}\frac{N+8}{6}-6u\Delta\right]
\frac{1}{L^{d}}{\sum_k}'\frac{1}{(k^{2}+t)^{2}}\\
\Delta_{L}&=& -\left[u\Delta\frac{N+2}{3}-\frac{8+nN}{2}\Delta^{2}\right]
\frac{1}{L^{d}}{\sum_k}'\frac{1}{(k^{2}+t)^{2}},
\end{eqnarray}
\end{mathletters}
are the corresponding finite-size corrections. As one can see the
summand in Eq. (\ref{udre}) can be expressed as the first
derivative of the summand of Eq. (\ref{eqnp}) with respect to $t$.
So, at the fixed point, algebraic transformations similar to those
performed in the previous section lead to
\begin{mathletters}\label{udsc}
\begin{eqnarray}\label{udelta}
\tilde{u}^{\star}(n)L^{\epsilon}&=&u^{\star}(n)\left[1+\frac{1}{2}
(1+\ln y)\epsilon +2\epsilon F^{\prime}_{4,2}(y)\right]\\
\tilde{\Delta}^{\star}(n)L^{\epsilon}&=& \Delta^{\star}(n)\left[1+\frac{1}{2}
(1+\ln y)\epsilon + 2\epsilon F^{\prime}_{4,2}(y)\right],
\end{eqnarray}
\end{mathletters}
where the prime indicates that we have derivative of the function
$F_{4,2}(y)$ with respect to its argument.

The results for the disordered system simply follow by setting $n=0$.
From the results for the shift of the critical temperature
(\ref{svqst}) and the
renormalization of the coupling constant $u$, given by Eq.
(\ref{udelta}), we reproduce the results
for the pure FSS case, by setting $\Delta=0$ and $n=0$. Moreover,
this result still holds even if we find the FSS corrections
after the analytical continuation to $n=0$, expressing the
commutativity of the problem.
\section{Verification of FSS}\label{secfss}

Let us consider the scaling variables
\begin{equation}
\mu(n)=\tilde{t}(n)L^{d/2}/\sqrt{\tilde{u}(n)},\qquad
\lambda(n)=\tilde{\Delta}(n)/\tilde{u}(n).
\end{equation}
At the fixed point they can be expressed in terms of scaling variable
$y=tL^{1/\nu(n)}$:
\begin{eqnarray}\label{muzvg}
\mu^{\star}(n)&=&\frac1{\sqrt{u^\star(n)}}\left\{y-\frac{1}{4}y\left[1
+ \frac{(4-N)(4-nN)}{16(N-1)-nN(N+8)}\ln y\right]\epsilon \right.\nonumber\\
&&- \left.\frac{12N(n-1)}{16(N-1)-nN(N+8)}F_{4,2}(y)\epsilon -
 yF^{\prime}_{4,2}(y)\epsilon\right\}
\end{eqnarray}
and
\begin{equation}\label{lamzg}
\lambda^{\star}(n)=\frac{4-N}{3(4-nN)}.
\end{equation}
In the limit $n=0$, Eqs. (\ref{muzvg}) and (\ref{lamzg}) yield the
following scaling variables describing the disordered system
(\ref{eqGLW}):
\begin{eqnarray}
\label{muzv}
\mu^{\star}:=\mu^{\star}(0)&=&\frac1{\sqrt{u^\star(0)}}
\left\{y-\frac{1}{4}y\left[1
+ \frac{4-N}{4(N-1)}\ln y \right]\epsilon \right.\nonumber\\ &+&
\left.\frac{3N}{4(N-1)}F_{4,2}(y)\epsilon -
 yF^{\prime}_{4,2}(y)\epsilon\right\},
\end{eqnarray}
where $y=tL^{1/\nu_{R}}$, and
\begin{equation}\label{lamz}
\lambda^{\star}:=\lambda^{\star}(0)=\frac{4-N}{12}.
\end{equation}
These equations verify the finite-size scaling hypotheses and show
that we are really dealing with a one-variable problem, since the
second variable $\lambda^{\star}$ is a fixed universal number. At
the critical point $t=0$, since the constant $F_{4,2}(0)=-8\ln2$,
see Ref. \cite{CT}, we have
\begin{equation}
\label{hmu}
\mu_0^{\star}:=\mu^{\star}|_{t=0}=
-\frac{N\ln2}\pi\sqrt{\frac{3\epsilon}{N-1}}.
\end{equation}
Numerical values for different thermodynamic quantities can be
obtained with the help of Eqs. (\ref{lamz}) and (\ref{hmu}). Note
that the scaling variable $\mu_0^{\star}$ is proportional to
$\sqrt{\epsilon}$. Consequently all the $\epsilon$-expansion
results will be expressed in power of $\sqrt{\epsilon}$ as it was
the case for the pure systems (see \cite{BZJ} for example).

\section{Cumulants and self-averaging}\label{secsa}

In Ref. \cite{BFMMPR}, the Binder cumulant $B$ and the
the relative variance $R$ (Eqs. (\ref{Bc}) and (\ref{rnc})) are calculated
analytically and numerically at the critical point $T=T_c$
in the asymptotic regime $\lambda \simeq 1/4$
for $N=1$ and $d=4$. In ref.\cite{Bet} (see also
\cite{WeD}), the same quantities were
calculated numerically for $N=1$ and $d=3$. In both cases the results
showing that the system exhibits a lack of SA.

In the remainder of this section we concentrate on the calculation
of the cumulants $B$ and $R$ (\ref{Bc}) and (\ref{rnc}) in cases
$d\geq4$ and $d=4-\epsilon$. The meaning to consider the case $d\geq4$
is in its simple analytical nonperturbative treatment. Although
the results based on the $\epsilon$ - expansion give only a
qualitative description  of the three dimensional physics, we hope
that they shed some light at least on the applicability of the
theory for studying diluted models.

Let us first note that if $1<N<4$ and $d=4-\epsilon$, the case
under consideration applies to the situation (see Eq. (\ref{nur}))
where $\nu_{R}>2/d$ and randomness is relevant (R-case). Up to the
first order in $\epsilon$, due to the RG arguments, no SA must be
expected near the critical point \cite{ABH}. This statement is
supported also by our RG calculations. In Table \ref{table1} and
Table \ref{table2} we
present the corresponding universal numbers for B and R at $d\geq
4$ and $d=3$ in the region $Lt^{\nu_{R}}=\frac{L}{\xi}\ll1$, i.e.
in the vicinity of the critical point. The calculations are
performed with variable $\mu=0$ for $d \geq 4$ and
$\mu=\mu_0^{\star}$ from Eq. (\ref{hmu}) (setting $\epsilon=1$)
for $d=3$, and with variable $\lambda$ taken from Eq. (\ref{lamz})
in both cases. The asymptotic behavior for small $\mu$ is presented in
the Appendix. The numerical values of B and R in the random case
and for $N=1$, presented in Table \ref{table1}, are in full agreement with
those obtained in Ref. \cite{BFMMPR}, while those of B for
the pure case and $N=1$ (Table I and Table \ref{table2}) are
in full agreement with Ref. \cite{BZJ}.

The random case $N=1$ for $d<4$ can not be considered within the
present expansion,
because of the apparent divergence of $\mu_0^{\star}$ that takes
place to the used order in $\epsilon$. Up to now there are only
numerical values $B=\frac{2}{3}$,  $g_{4}=0.448$ and
$R=g_{2}=0.150(7)$ obtained in \cite{Bet} through Monte Carlo
simulations. What is possible to calculate here are the
corresponding values of B and R very close to $N=1$, e.g. for
$N=1.001$. For completeness these results are presented in Table
II. More general, one can see that if $N \to 1$, then $B \to
\frac{2}{3}$ and $R\to 0$, i.e. the system exhibits SA. This
evident discrepancy with the reality is due to the wrong
assumption that some information about the random case $N=1$ can
be obtained from the above formulas in this limiting case. As it
was pointed out the correct treatment of the case $N=1$ seems to
be a more difficult computational problem.

The finite size correction to the bulk critical behavior of the
cumulants $B$ and $R$ in the region
$Lt^{\nu_{R}}=\frac{L}{\xi}\gg1$, i.e. away from the critical point,
are obtained with the help of the
asymptotics $\mu\gg1$, given in the
Appendix (Eqs. \ref{AM2}, \ref{AM4}). According to the analysis
presented there, we obtain for Binder's cumulant
\begin{equation}\label{Bdkt}
B=1-\frac13\left(1+\frac{2}{N}\right)
\left[1+\frac{3\lambda-1}{3\mu^2}\right]
+{\cal O}\left(\frac{1}{\mu^{3}}\right).
\end{equation}
For the cumulant $R$ we get
\begin{equation}\label{relative}
R=\frac{\lambda}{\mu^2}+{\cal O}\left(\frac{1}{\mu^{3}}\right).
\end{equation}

The final results can be obtained by replacing $\lambda$ and $\mu$
by their respective expressions evaluated at the fixed points of
the model given in Eqs. (\ref{muzv}) and (\ref{lamz}). So, to the
lowest order in $\epsilon$, we have
\begin{equation}\label{Biz}
B=1-\frac13\left(1+\frac{2}{N}\right) + {\cal
O}\left(\frac{\xi}{L}\right).
\end{equation}
and
\begin{equation}
\label{Riz}
R=\frac{4-N}{8(N-1)}\epsilon\left(\frac{\xi}{L}\right)^4+{\cal
O}\left(\frac{\xi^{5}}{L^{5}}\right).
\end{equation}

It is interesting to compare Eq. (\ref{Bdkt}) with the
corresponding result for the pure system \cite{ChD}
\begin{equation}
B_{pure}=1-\frac{1}{3}\left(1+\frac{2}{N}\right)
\left(1-\frac1{3\mu_p^2}+{\cal
O}\left(\frac1{\mu_p^4}\right)\right).
\end{equation}
Up to the lowest order in $\frac{1}{\mu}$ they coincide for
$1<N<4$, moreover for $N=4-\delta$ (with $\delta\ll1$), we have
$\mu=\mu_p+{\cal O}(\epsilon\delta)$ and $B=B_{\rm pure}+{\cal
O}(\epsilon\delta)$. The result (\ref{Riz}) confirms the statement
\cite{ABH} that away from the critical point, a strong SA emerges
in the system, as $R \rightarrow 0$ with $\frac{L}{\xi}\gg1$.


\section{Conclusions}\label{discussion}

In the present paper we propose a general scheme for the FSS scaling
analysis of a finite disordered ${\cal O}(N)$ system. The method, we use
here, is an extension of the field theoretical methods used to analyse
FSS properties in pure systems. The nature of the symmetry (obtained as a
consequence of the use of the replica trick, which removes the disorder)
of the model complicates the perturbative structure of the theory in
comparison with the corresponding ${\cal O}(N)$ pure one.
Remind that, the final results for the disordered system are obtained
by making the number of replicas vanishing.
Our results concern mainly systems with number of components larger
than 1 i.e. non Ising systems. Their extension to
Ising systems requires higher loop calculations, because of the
degeneracy of the one loop order RG equations.


Our main results are related to the formulation of the problem for
some number of components $N$ of the fluctuating field for dimensions
$d>4$ and $d=4-\epsilon$. Due to the presence of randomness, it is
shown that we are dealing with two variables problem with scaling
variables $\mu=tL^{d/2}u^{-1/2}$ and $\lambda=\Delta/u$. In the mean
field regime $d>4$ our results are a generalization for $N>1$ of those
obtained in~\cite{BFMMPR} for $N=1$. Evaluating numerically the
corresponding analytic expressions for the Binder's cumulant $B$ and
the relative variance $R$, we demonstrate a monotonic increase of $B$ as a
function of $N$ in both pure and random cases and a monotonic decrease
of $R$ (to zero for $N=4$) in the random case (see Table
\ref{table1}).

The $\epsilon$-expansion to first order in $\epsilon$, shows that
close to the critical temperature, one can express the physical
observables, the shift of the critical temperature and the renormalization
of the coupling constants in terms of $\mu$ and $\lambda$. In the
random fixed point the parameter $\lambda$ takes an universal value (see
equation (\ref{lamz})). It is found that the distance, over which the
bulk critical temperature is shifted, is proportional to
$L^{-1/\nu_R}$, in agreement with the statement of Ref.~\cite{ABH}.
This result, combined with fenomenological FSS, gives rise to the lack of
SA. The scaling parameters $\mu$ and $\lambda$, also enter in the
final expression for the Binder cumulant (\ref{Bc}) and the relative variances
(\ref{rnc}), giving explicit expressions for them in the different asymptotic
regimes. The numerical calculation of the above parameters permits also
the verification of the FSS and SA (see Table \ref{table2}). The last
is shown to be absent in the regime $\nu_R > 2/d$, where randomness are
relevant and our analysis explicitly shows that in this regime the relative
variance is always non-zero for the case of stability of the RG solutions,
i.e. $1<N<4$ and $d=4-\epsilon$.

One can also try to repeat the
analysis without the use of the replica-trick.
For this aim, one needs to define in a proper way the procedure in the
zero-mode approximation, when rescaling the random part of the Hamiltonian
for finite system. One can realize that this is the ${\bbox k}=0$ part of
this term, which will
give a contribution to the free-energy. This scheme permits formally to end
with the same expressions for the shift of the critical temperature and the
renormalization constants, as we did within the replica formalism.

In our opinion, the present FSS study can be also applied in the ``canonical''
case \cite{ABW}, where the disorder is characterized by a constant
total number of the occupied sites (or bonds), instead of the constant
average density.
We hope that results, similar to the bulk case, will also hold in the case
of finite geometry, relating in this way our theoretical findings with the
Monte-Carlo simulations.

\acknowledgments
The authors aknowledge the hospitality of the Abdus Salam International
Centre for Theoretical Physics, Trieste, where part of this work was
written. E.K. and N.S.T. acknowledge also the financial support of
Associate \& Federation Schemes of ICTP. E.K. is also supported by the
Spanish DGES Contract No. PB97-0076.

\appendix
\section*{Finite-size scaling behavior of the even moments of the order
parameter}

In this Appendix we present the mathematical details of how to obtain
the asymptotic of the averages:
\begin{mathletters}\label{averages}
\begin{equation}\label{averages1}
\overline{{\cal M}_{2m}}=
\frac{(uL^d)^{-m/2}}{\sqrt{2\pi}}\int_{-\infty}^\infty dx\frac{{\cal
I}_{N+2m}(\mu+\sqrt{\lambda} x)} {{\cal I}_{N}(\mu+\sqrt{\lambda}
x)}e^{-x^2/2},
\end{equation}
\begin{equation}
\overline{({\cal
M}_2)^2}=\frac{(uL^d)^{-1}}{\sqrt{2\pi}}\int_{-\infty}^\infty
dx\left[\frac{{\cal I}_{N+2}(\mu+\sqrt{\lambda} x)}{{\cal
I}_{N}(\mu+\sqrt{\lambda} x)}\right]^2e^{-x^2/2},
\end{equation}
\end{mathletters}
where we have introduced the function
\begin{equation}\label{fI}
{\cal I}_{N}(z)=S_N
\int_{0}^{\infty}d\Phi \Phi^{(N-1)}\exp\left\{-\frac12\left[z
\Phi^{2}+\frac1{12}\Phi^{4}\right]\right\}.
\end{equation}
The integral in the definition of function ${\cal I}_{N}(z)$ given
by Eq. (\ref{fI}), may be evaluated in terms of parabolic cylinder
functions $D_{p}(z)$ using the identity \cite{gradsteyn65}
\begin{equation}
\int_0^\infty x^{\nu-1}e^{-\beta x^2-\gamma
x}dx=(2\beta)^{-\nu/2}\Gamma(\nu)
\exp\left({\frac{\gamma^2}{8\beta}}\right)
D_{-\nu}\left(\frac{\gamma}{\sqrt{2\beta}}\right).
\end{equation}
The result is:
\begin{equation}\label{NN}
{\cal I}_{N}(z)=(12\pi^2)^{N/4}
\exp\left(\frac{3z^{2}}{4}\right)D_{-N/2}(\sqrt{3}z).
\end{equation}

Now the above integrand in Eq. (\ref{averages1}) can be rewritten
in a very simple form
\begin{equation}\label{ratio}
{\cal M}_{2m}(x):=\frac{{\cal I}_{N+2m}(\mu+\sqrt{\lambda} x)}{{\cal
I}_{N}(\mu+\sqrt{\lambda} x)}=
\left(12\pi^2\right)^{\frac m2}
\frac{D_{-m-N/2}((\mu+\sqrt{\lambda}
x)\sqrt{3})} {D_{-N/2}((\mu+\sqrt{\lambda} x)\sqrt{3})}.
\end{equation}

For small $\mu\ll1$ (i.e. in the vicinity of the critical point)
the asymptotic form of the ratio (\ref{ratio}) is given by
\begin{eqnarray}\label{mull1}
{\cal M}_{2m}(x)&=&\left(12\pi^2\right)^{\frac m2}
\left\{\frac{D_{-m-N/2}(x\sqrt{3\lambda})}
{D_{-N/2}(x\sqrt{3\lambda})}\right.\nonumber\\
&&+\frac{\mu\sqrt3}2\left[N\frac{D_{-N/2-1}(x\sqrt{3\lambda})
D_{-m-N/2}(x\sqrt{3\lambda})}
{\left(D_{-N/2}(x\sqrt{3\lambda})\right)^2}
-(2m+N)\frac{D_{-m-N/2-1}(x\sqrt{3\lambda})}
{D_{-N/2}(x\sqrt{3\lambda})}\right]\nonumber\\ &&\left.+{\cal
O}(\mu)^2\right\}.
\end{eqnarray}

In the mean-field regime and at the critical point we have $\mu=0$ and
${\cal M}_{2m}$ is equal to the first term in the r.h.s. of Eq.
(\ref{mull1}).

For large $\mu\gg1$, the asymptotic behavior of the ratio ${\cal
M}_{2m}$ is obtained with the help of the well known Watson's Lemma:

\vspace{1cm}

\noindent {\bf Lemma:} (see for example \cite{lemmaf})

{\it
Suppose $\alpha>0$, $\beta>0$ and $f(x)$ an analytic function in a
neighborhood of $x=0$,
$$
f(x)=\sum_{k=0}^\infty \frac{f^{(k)}(0)}{k!} x^k, \ \ \ |x| < R,
$$
and that
$$
|f(x)|\leq c_1e^{c_2x^\alpha}, \ \ \ x\in[R,X],
$$
for positive constants $c_1$, $c_2$. Then:
$$
\int_0^Xx^{\beta-1}e^{-s x^\alpha}f(x)dx\sim\frac1\alpha
\sum_{k=0}^\infty s^{-(k+\beta)/\alpha}
\Gamma\left(\frac{k+\beta}\alpha\right)\frac{f^{(k)}(0)}{k!},
$$
as $s\to\infty$ in the sector $|\arg s|<\frac\pi 2$. }

\vspace{1cm}


According to this Lemma, from Eqs. (\ref{fI}) (with 
$z=\mu+x\sqrt\lambda$) and (\ref{ratio}), we have
\begin{equation}\label{AM2}
{\cal M}_2(x)=\left(12\pi^2\right)^{1/2}\frac N\mu
\left[1-\frac{x\sqrt\lambda}\mu+
\frac{6x^2\lambda-N-2}{6\mu^2}+
{\cal O}\left(\frac1{\mu^3}\right)\right]
\end{equation}
and
\begin{equation}\label{AM4}
{\cal M}_4(x)=
12\pi^2\frac{N(N+2)}{\mu^2}
\left[1-\frac{2x\sqrt\lambda}{\mu}+
\frac{9x^2\lambda-N-3}{3\mu^2}
+{\cal O}\left(\frac1{\mu^3}\right)\right].
\end{equation}

Using the asymptotic of ${\cal M}_{2}$ and ${\cal M}_{4}$ for large
$\mu$, we can get the behavior of the cumulants $R$ and $B$ in both
cases $d \geq 4$ and $d=4-\epsilon$. They are given by
\begin{equation}
B=1-\frac13\left(1+\frac{2}{N}\right)
\left[1+\frac{3\lambda-1}{3\mu^2}\right]
+{\cal O}\left(\frac{1}{\mu^{3}}\right).
\end{equation}
For the cumulant $R$ we get
\begin{equation}
R=\frac{\lambda}{\mu^2}+{\cal O}\left(\frac{1}{\mu^{3}}\right).
\end{equation}


\begin{table}

\caption{Numerical values for the Binder cumulant $B$ from
Eq. (\ref{Bc}) and the relative variance $R$ from Eq.
(\ref{rnc}) in the mean-field regime i.e. $d\geq4$.}
\begin{tabular}{ccccc}
&\multicolumn{2}{c}{Random}&\multicolumn{2}{c}{Pure}\\
N       & B         & R          & B          & R          \\
\tableline
1       & 0.216368  & 0.310240   & 0.270520   & 0          \\
2       & 0.451486  & 0.111381   & 0.476401   & 0          \\
3       & 0.533513  & 0.038365   & 0.543053   & 0          \\
4       & 0.575587  & 0          & 0.575587   & 0          \\
\end{tabular}
\label{table1}
\end{table}

\begin{table}
\caption{Numerical values for the Binder cumulant $B$ from
Eq. (\ref{Bc}) and the relative variance $R$ from Eq. (\ref{rnc}) at
$d=3$}

\begin{tabular}{ccccc}
&\multicolumn{2}{c}{Random}&\multicolumn{2}{c}{Pure}\\
N       & B         & R          & B          & R          \\
\tableline
1       & -         & -          & 0.400024   & 0          \\
1.001   & 0.666334  & 0.000427   & 0.400328   & 0          \\
2       & 0.602793  & 0.061279   & 0.547496   & 0          \\
3       & 0.625783  & 0.022688   & 0.592813   & 0          \\
4       & 0.640628  & 0          & 0.614002   & 0          \\
\end{tabular}
\label{table2}
\end{table}

\begin{thebibliography}{99}
\bibitem{eqv} The technique that involves application of the RG
transformations to the disordered system directly \cite{L} gives
the same fixed points as the technique that involves first
removing the disorder by averaging and then employing the RG
transformation \cite{GL}. For a direct nonperturbative derivation
of this equivalence see \cite{Em}.
\bibitem{footnote} The indices ``r'' and ``p'' will refer to the
``random'' and the effective ``pure'' system.
\end{thebibliography}
\end{document}